# Amorphous phase-change memory alloy with no resistance drift

Xiaozhe Wang[1,#], Ruobing Wang[2,#], Suyang Sun[1,#], Ding Xu[1], Chao Nie[1], Zhou Zhou[1], Chenyu Wen[1], Junying Zhang[1], Ruixuan Chu[1], Xueyang Shen[1], Wen Zhou[1], Zhitang Song[2], Jiang-Jing Wang[1,*], En Ma[1,*], Wei Zhang[1*]

[1]Center for Alloy Innovation and Design (CAID), State Key Laboratory for Mechanical Behavior of Materials, Xi'an Jiaotong University, Xi'an, 710049, China.
[2]National Key Laboratory of Materials for Integrated Circuits, Shanghai Institute of Microsystem and Information Technology, Chinese Academy of Sciences, Shanghai 200250, China

[#]These authors contributed equally to this work.

*Emails: j.wang@xjtu.edu.cn, maen@xjtu.edu.cn, wzhang0@mail.xjtu.edu.cn

**Abstract:**
Spontaneous structural relaxation is intrinsic to glassy materials due to their metastable nature. For phase-change materials (PCMs), the resultant temporal change in electrical resistance seriously hamper neuromorphic computing applications. Here, we report an ab-initio-calculation-informed design of amorphous PCM composed of robust "molecule-like" motifs, depriving the amorphous alloy of critical structural ingredients responsible for the relaxation and hence the resistance drift. We demonstrate amorphous $CrTe_3$ thin films that display practically no resistance drift at any working temperature from −200 to 165 °C, and highlight the multilevel encoding ability via a hybrid opto-electronic approach. We further reveal that the same no-drift behavior holds for melt-quenched amorphous $CrTe_3$ in electronic devices. Moreover, the application potential of $CrTe_3$ is testified by its incorporation in a vehicle with automatic path-tracking function. Our work opens a new avenue to achieving requisite properties for potential phase-change neuromorphic computing via judicious design of disordered PCM alloys.

Data-intensive applications, such as Artificial Intelligence (AI) and the Internet of Things (IoT), are having a profound impact on nearly all aspects of life. The huge amount of data being generated at an ever-increasing rate poses formidable challenges to efficient data storage and processing. Neuromorphic computing or in-memory computing (IMC) based on resistive-switching nonvolatile memory (NVM) arrays, unifying storage with computing at the same location, can substantially improve the computing efficiency at a much lower energy cost [1-4]. The key to high-storage-density neuromorphic computing is to obtain as many distinguishable resistance levels as possible in every memory/computing cell [5]. Yet, high-precision IMC is facing a bottleneck problem associated with the intrinsic randomness of materials upon programming [6], especially when operated at a working temperature (*T*) different from room temperature (RT). Continuous cooling is feasible at power farms, but for the power-limited edge computing platforms, temperature effects [7-9] on accurate multilevel programming are much more critical.

Chalcogenide phase-change materials (PCMs), which utilize the large resistance contrast between their amorphous and crystalline states for data encoding [10,11], are one of the leading candidates for NVM. The flagship $Ge_2Sb_2Te_5$ (GST) alloys with moderate doping [12] or heavy alloying [13] have enabled industrial production of high-density 3D Cross Point memory cards for computing servers and data centers with an operating *T* of 60~80 ºC, as well as embedded memory units for automotive with a working *T* of −40~165 ºC [14]. Very recently, it has also been shown that crystalline and amorphous PCM can survive the harsh space environment outside of the International Space Station [15], where temperature can cycle between −120 and 120 °C 16 times per day. Beyond the success in these binary storage applications, PCMs do show multilevel storage capacity. But the spontaneous structural relaxation – aging – of amorphous PCM causes intolerable resistance drift [16-18]. This is a pressing problem threatening the device accuracy needed for neuromorphic computing applications [19].

Massive research efforts have been devoted to overcoming the adverse impact of resistance drift. On the device side, several approaches using mixed-precision computing [19], algorithm compensation [20], metal liner confinement [21] and multi-PCM units [22] were developed to partially mitigate the drift problem. It remains unclear whether these workaround solutions can still function at elevated temperatures or upon extreme temperature cycling, but in any case, they add complexities to high-density device manufacturing and accurate multilevel programming. Therefore, a fundamental solution on the materials side, i.e., PCMs with intrinsically low drift and randomness, are sorely needed. Previous attempts employing impurity doping, heterostructure confinement and down-scaling the film thickness only managed to reduce the drift tendency at RT [23-29]. However, there is so far no candidate PCM that can generate thousands of robust resistance states to function across a wide range of operating temperatures for various application scenarios. In the following, we offer a materials solution to diminish relaxation/drift once and for all, via a paradigm shift towards "molecular-glass-like" amorphous PCMs (a-PCMs). This approach enabled us to discover an a-PCM that is intrinsically insusceptible to resistance drift at all working temperatures practically encountered in IMC operations.

**Design of "molecular-glass-like" a-PCM**

Upon aging, a-PCMs are subject to structural relaxation. Conventional a-PCMs experience a gradual (i) reinforcement of Peierls distortion of (defective) octahedral motifs [30] and (ii) disappearance with increasing time (*t*) of salient structural defects[30-32], including "wrong" bonds (e.g. Ge−Ge homopolar bonds) and tetrahedral motifs. Both of these two factors result in resistance drift. Our key idea is then to find an a-PCM that intrinsically has no such tendencies during the time- and temperature-induced relaxation. We first noticed that Peierls distortion is nearly non-existent in all crystalline polymorphs of 3*d* transition metal (TM) tellurides at Te-rich compositions (Te:TM ≥ 2:1), where each 3*d* transition metal atom is coordinated by six Te atoms, forming an octahedral motif with symmetric bonding along each direction. Although the octahedral motifs can be slightly distorted, with the Te-TM-Te bond angle being slightly smaller or larger than 90°, there is no long/short bond correlation to establish *bona fide* Peierls distortion (see Supplementary Fig. S1). We conjectured that this scenario could be retained to the amorphous counterpart of such tellurides. Given the Te-rich composition, the probability to form wrong (TM-TM) bonds in the amorphous state is also reduced. It is then promising that the two factors (i) and (ii) in the preceding paragraph can both be minimized.

To this end, we scrutinized all 3*d* transition metal tellurides. Among them, $CrTe_3$ is the richest in Te composition. Fig. 1a shows the relaxed crystalline (c-) $CrTe_3$ calculated via spin-polarized density functional theory (DFT). First, the directional bonds of c-$CrTe_3$ show no obvious Peierls distortion; the bond lengths are nearly identical, and the maximum difference is as small as 0.03 Å, an order of magnitude smaller than that of rhombohedral GeTe, ~0.4 Å. Second, the crystal shows a layered structure composed exclusively of corner- and edge-sharing [$CrTe_6$] octahedra. It is thus well possible that its amorphous phase (a-$CrTe_3$) would also contain a very high density of non-defective [$CrTe_6$] octahedra with little Peierls distortion. To see if this hypothesis pans out, we carried out DFT-based ab initio molecular dynamics (AIMD) calculations to obtain a-$CrTe_3$ models of 200 atoms following a standard melt-quenched approach. We first used the computed crystalline density directly, and the obtained atomic structure is shown in Fig. 1a. As highlighted using coordination polyhedra, almost all Cr atoms indeed form [$CrTe_6$] octahedra (48 out of 50 Cr atoms, and the rest 2 Cr octahedra have only one missing neighbor). This observation is further confirmed by quantitative structural analyses shown in Supplementary Fig. S2a.

We next evaluated the degree of Peierls distortion in the amorphous counterpart, i.e., a-$CrTe_3$ model, by calculating the angular limited three-body correlation (ALTBC) at RT (~25 ºC) over directional bonding pairs (bond angle ~155−180º). In stark contrast with conventional PCM with an extended wing-shape [30], a-$CrTe_3$ displays a single primary peak around 2.72 Å in the ALTBC profile at RT (Fig. 1b), which matches that of the c-$CrTe_3$ (marked by black circle). We also calculated the ALTBC profiles of a-$CrTe_3$ at other (lower and higher than RT) temperatures (Supplementary Fig. S2b), and they show a picture consistent with that at RT. Further optimization of the amorphous model at RT leads to a volume expansion by ~2.1%, but the structural features remain the same (see Supplementary Fig. S2c). Four additional a-$CrTe_3$ models in both crystalline and amorphous density were calculated, which confirm that a-$CrTe_3$ is consisted of [$CrTe_6$] octahedra connected in a disordered fashion, without obvious Peierls distortion and structural defects. In fact, octahedra are the only local configuration observed, and these motifs can hence be deemed “molecule-like”. We prolonged the cooling time by a factor of two and three in forming a-$CrTe_3$, and this structural feature remained the same (Supplementary Fig. S3).

With the desired structural features in hand, the next challenge is the requisite electronic properties. For IMC applications, it is imperative for the PCM to have a large contrast window in electrical resistance upon switching. The computed electronic density of states (DOS) does predict a major difference in electronic structure between c-$CrTe_3$ and a-$CrTe_3$ (Fig. 1c). The crystalline phase is a semiconductor stabilized by antiferromagnetic interactions [33]. For the amorphous phase, we considered a couple of magnetic configurations, and the band gap is always filled. Fig. 1c shows the DOS for the ferromagnetic state, and other magnetic configurations are included in Supplementary Fig. S4. The open versus filled band gap should result in a sizable difference in electrical resistivity between the two states, with the amorphous state being the lower-resistance state, which was observed in a similar layered PCM, $CrGeTe_3$ [34,35]. Importantly, as noted in Supplementary Fig. S2, the same octahedral configuration ("molecules" as well as their number density) stays predominant even when the temperature is elevated to 165 °C, suggesting that the a-PCM should also have an excellent thermal stability across a wide range of operating temperatures.

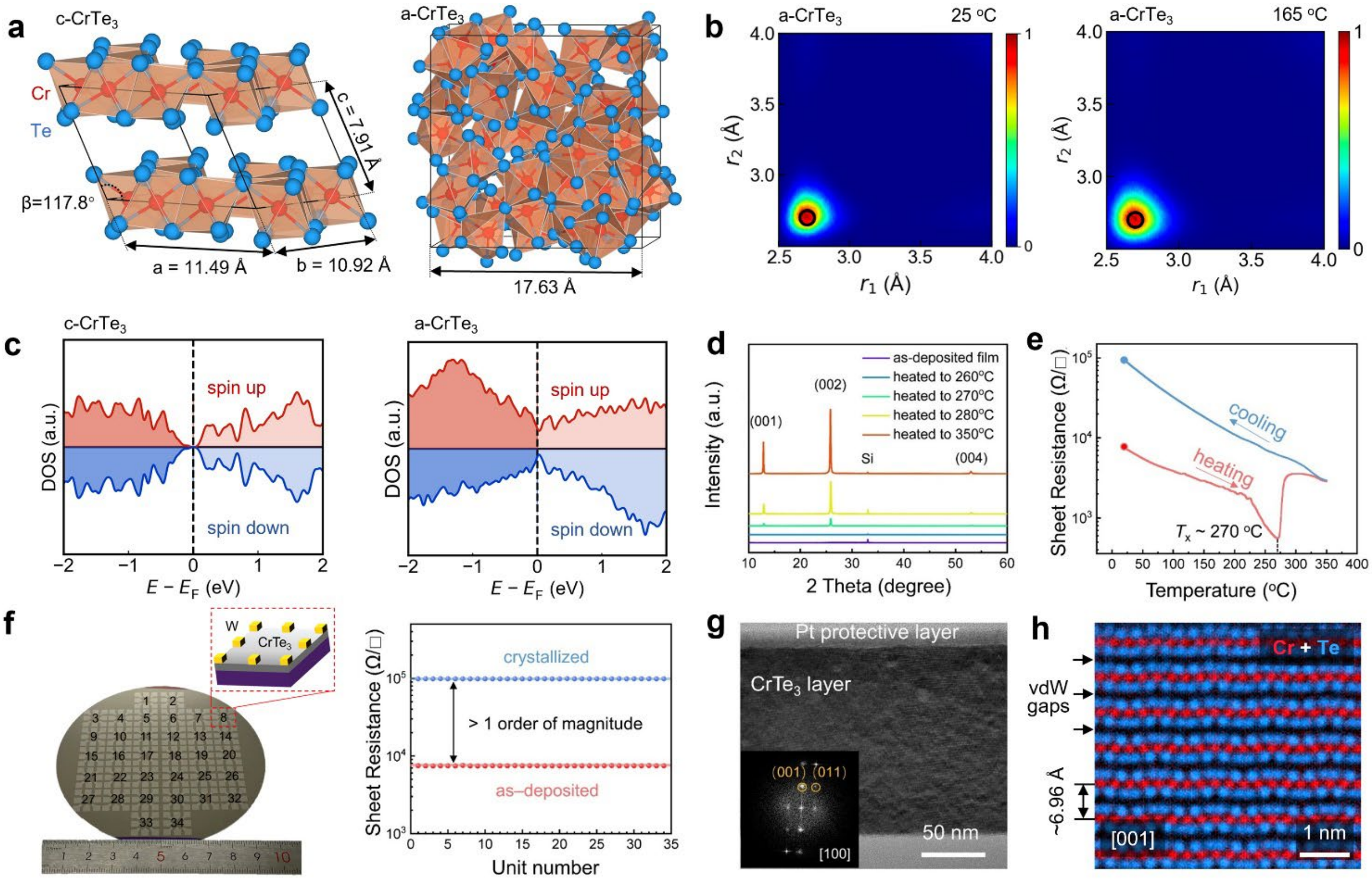

**Fig. 1 | Materials design, synthesis and characterizations.** (**a**) The relaxed atomic structures of c-$CrTe_3$ and melt-quenched a-$CrTe_3$. The Cr and Te atoms are rendered with red and light blue spheres and the [$CrTe_6$] octahedra are highlighted by orange polyhedra. (**b**) The ALTBC plot of a-$CrTe_3$ showing no Peierls distortion at ~25 °C and ~165 °C. The black circle indicates equal bonds in c-$CrTe_3$ (~2.7 Å). (**c**) The calculated DOS of c- and a-$CrTe_3$. (**d**) The XRD patterns of the as-deposited thin film and the thin films heated to 260 °C, 270 °C, 280 °C and 350 °C, respectively. (**e**) The temperature dependence of sheet resistance of $CrTe_3$ film heated up to 350 °C with a heating rate of 10 °C /min, and was then cooled to RT. (**f**) The van der Paul measurement of 34 units on a 4-inch wafer. The 0.5 × 0.5 $cm^2$ $CrTe_3$ area in the middle of the substrate is contacted by 8 tungsten electrodes (marked in yellow), which were patterned at the edges of the 1.2 × 1.2 $cm^2$ substrate. (**g**) The TEM bright-field image and (**h**) the atomic-scale elemental mapping of c-$CrTe_3$. The positions of the Te and Cr columns are marked in blue and red, respectively.

**Wafer-scale synthesis of high-quality $CrTe_3$ films**

We deposited $CrTe_3$ thin films of ~50−150 nm in thickness on 4-inch-diameter $SiO_2$/Si substrates at RT, using a pure Cr target and a pure Te target via magnetron sputtering. A ~10 nm thick ZnS-SiO2 capping layer was grown on top of the $CrTe_3$ film inside the vacuum chamber to prevent oxidation and evaporation. Then we heated several ~150 nm as-deposited thin films up to 260 °C, 270 °C, 280 °C and 350 °C, respectively, with a heating rate of 10 °C /min. For the as-deposited film and the one heated to 260 °C, X-ray diffraction (XRD) patterns at RT (Fig. 1d) show no crystalline diffraction peaks, except for the silicon substrate. For higher annealing temperatures, c-$CrTe_3$ peaks of (100), (200) and (400) appeared, indicating that these thin films formed an out-of-plane texture. The lattice parameters obtained by XRD are 11.53, 11.19 and 7.79 Å, in good agreement with DFT calculation, 11.49, 10.92 and 7.91 Å. From the relatively sharp XRD peaks, the size of grains appears to be at least several hundreds of nanometers. As estimated from X-ray reflectometry (XRR), the mass density of the as-deposited amorphous film was ~6.12 $g/cm^{-3}$, and was increased to ~6.39 $g/cm^{-3}$ upon crystallization (Extended Data Fig. 1a), corresponding to a density increase ($\Delta\rho/\rho_{crystal}$) of ~3.9%, which is smaller than that of GST, ~6.4% [36].

Next, we measured the electrical resistance of the as-deposited thin film upon *in situ* heating to 350 °C with a heating rate of 10 °C / min using the van der Pauw method in an Argon-protected environment. In agreement with our DFT prediction, Fig. 1e shows an inverse resistance contrast upon crystallization for $CrTe_3$, opposite to GST. The sheet resistance of the as-deposited amorphous film was ~7 kΩ/□ at RT, slowly and gradually decreasing upon heating. Only after ~230 °C, did the sheet resistance drop more quickly (Supplementary Fig. S5), reaching a minimum at $T_x$ ~270 °C, where crystallization started, which is consistent with the XRD data. The sheet resistance of the crystallized film was measured to be 100 kΩ/□ at RT. According to the Hall effect measurements, the carrier concentration and mobility of the as-deposited film were $5.1 \times 10^{20}$ $cm^{-3}$ and $6.8 \times 10^{-1}$ $cm^2$ $V^{-1}$ $s^{-1}$, respectively, which were reduced to $6.6 \times 10^{19}$ $cm^{-3}$ and $2.8 \times 10^{-1}$ $cm^2$ $V^{-1}$ $s^{-1}$) in the crystallized film. To assess the reproducibility, we fabricated 34 tungsten-electrode-based test cells, each containing eight electrodes, and measured their sheet resistance of the as-deposited amorphous phase and the fully crystallized phase at 350 °C. All the 34 test cells gave highly consistent resistance values (Fig. 1f).

We carried out transmission electron microscopy (TEM) experiments to gain a microscopic view of the completely crystallized phase. As shown in Fig. 1g, the cross-sectional bright-field image looks homogeneous with limited intensity contrast, and no grain boundary can be observed at this magnification. The Fast Fourier Transform pattern (FFT) confirms the crystal orientation of the c-$CrTe_3$ film to be [001]. The zoom-in TEM images taken at locations from the top to the bottom of the film consistently show well-aligned (001) lattice planes parallel to the substrate surface (Extended Data Fig. 1b), consistent with the XRD results, indicating epitaxial-growth-like crystal quality even though no seeding layer was used in the deposition. The scanning transmission electron microscopy – high-angle annular dark field (STEM-HAADF) image and the corresponding elemental mapping images show a clear arrangement of slabs of $CrTe_3$ trilayers sandwiching van der Waals (vdW) gaps, see Fig. 1h and Extended Data Fig. 1c. We thoroughly checked the structural features at different locations, confirming the highly textured large grains across the 4-inch c-$CrTe_3$ thin film (Extended Data Fig. 1c). The electron-transparent region in the TEM foil (lateral dimension >5 μm) of the 50-nm c-$CrTe_3$ thin film (Extended Data Fig. 1d) is nearly a single grain. This structural uniformity leads to highly consistent resistance values across the whole silicon wafer (Fig. 1f).

**Suppressed resistance drift at extreme temperatures**

The resistance drift behavior of amorphous PCMs is usually characterized using $R(t) = R_0 (t / t_0)^{\nu}$ ($\sigma(t) = \sigma_0 (t / t_0)^{-\nu}$), where $R_0$ ($\sigma_0$) is the resistance (conductance) measured at the time $t_0$, and ν is the drift coefficient over time $t$. To avoid potential aging effects due to sample storage, we measured the resistance of a-$CrTe_3$ thin films immediately after their sputter preparation. We tested different film thickness from 50 to 150 nm, which all showed an ultralow drift with ν ~0.001 at RT. With a drift coefficient of this level, resistance drift would have little impact on practical multilevel programming.

In what follows, we focus on the film of 50 nm, as such film thickness is more suitable for multilevel programming in our devices. Fig. 2a displays the as-deposited a-$CrTe_3$ thin film heated to increasingly elevated temperatures in a sequential order, and at each temperature the holding time was 1 hour. The a-$CrTe_3$ thin film showed a consistently low ν of 0.001−0.002 between RT and 150 °C. At 165 °C, its ν slightly increased to ~0.004. These values are far smaller than that of a-GST, ν ~0.11 between RT and 100 °C (Supplementary Fig. S6). Another issue of a-GST is that it crystallizes quickly at higher temperatures (e.g., when approaching 150 °C), making it unsuitable for embedded memory applications. Drift measurements were performed also for a-GeTe, a-$CrGeTe_3$ and a-$Sb_2Se_3$ thin films (Supplementary Fig. S6), which all showed significant resistance drift.

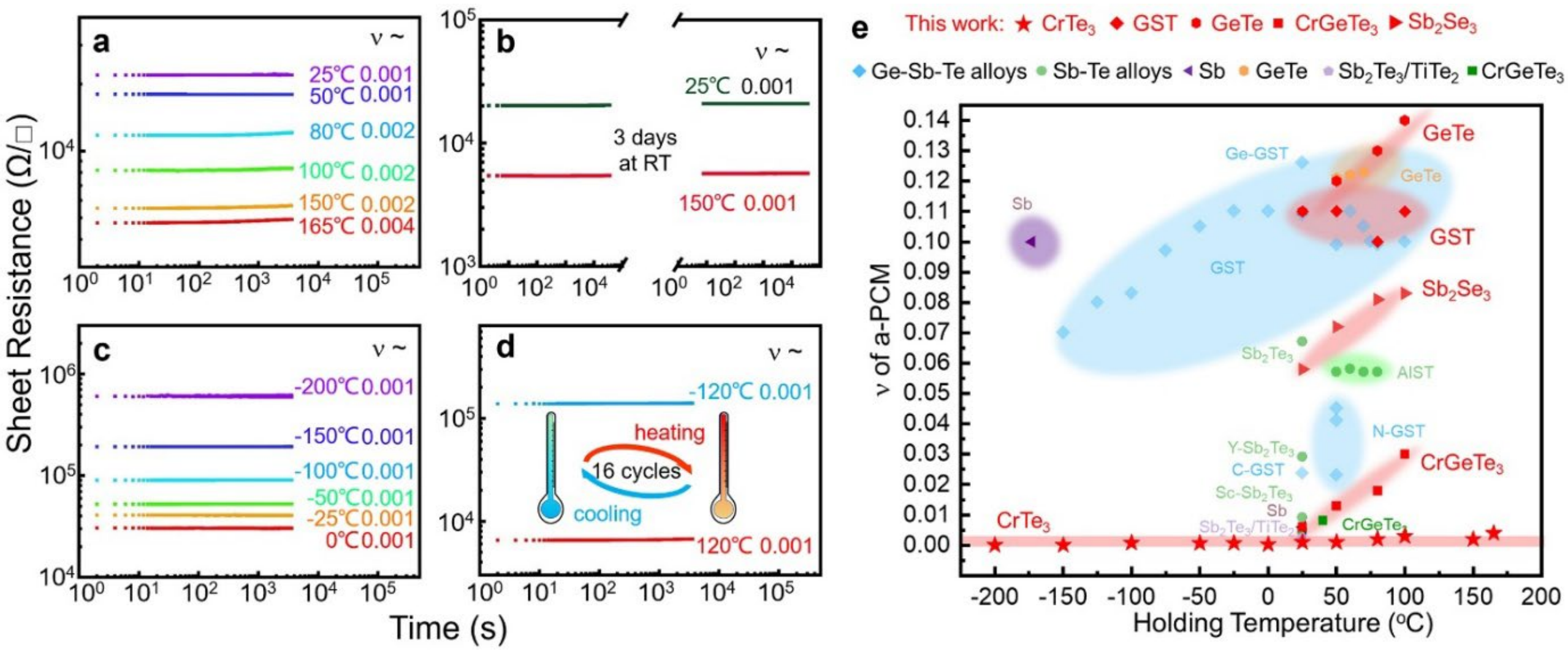

**Fig. 2 | Minimal resistance drift.** (**a**) The resistance drift measurement of as-deposited a-$CrTe_3$ films upon heating over 1 hour, each at a different holding temperature. (**b**) Two as-deposited a-$CrTe_3$ films were measured at 25 and 150 °C over 10 hours (first part, displayed on the left). After aging for 3 days at RT, they were measured at 25 and 150 °C over 10 hours again (second part, on the right). (**c**) The resistance of a-$CrTe_3$ film measured at 0 °C and below. (**d**) The a-$CrTe_3$ film was subjected to an extreme temperature cycling between –120 °C and 120 °C over 16 times in 24 hours, and subsequently its sheet resistance was respectively measured at –120 °C and 120 °C over time. The distinctive resistance levels were due to the varied thermally excited carrier concentration at different temperatures. (**e**) Summary of the measured drift coefficient of various amorphous PCM. The drift coefficient is highly temperature-dependent for other PCMs, e.g. ν of a-Sb can be enlarged from ~0.001 at RT to ~0.1 at −173 °C.

The ultralow drift behavior of a-$CrTe_3$ was also found to be robust over much longer time. We took two other fresh as-deposited thin films, and recorded their sheet resistance at 25 °C and 150 °C over 10 hours, respectively. Then the two films were stored for three days in air without Argon protection at RT. Subsequently, the two films were measured again at 25 °C and 150 °C over another 10 hours. Both sets of a-$CrTe_3$ films showed ν ~0.001 (Fig. 2b). It is noted that transition metal atoms are typically reactive, and they can still be oxidized despite the presence of the capping layers. But for $CrTe_3$, all Cr atoms form strong bonds with Te atoms in compact octahedra, presenting a good air stability.

We also tested the drift behavior of a-$CrTe_3$ at much lower temperatures. Our home-made setup was used to record the resistance at 0 and −25 °C, whereas a physical property measurement system (PPMS) utilizing the liquid helium refrigeration was employed to measure the resistance change at even lower temperatures. As shown in Fig. 2c, our a-$CrTe_3$ film again displayed very steady resistance with little drift, ν ~0.001, when held at any temperature from 0 ºC down to −200 ºC. We also confirmed that such minimal drift can be sustained upon extreme temperature cycling. We heated an as-deposited $CrTe_3$ thin film to 120 ºC, and then cooled it down to −120 ºC (heating/cooling rate of 5.4 °C/min). Such thermal cycling was performed for a total of 16 times within 24 hours. The resistance of the film was then measured at 120 ºC and −120 ºC, respectively. As shown in Fig. 2d, the ultralow drift remained unchanged, ν ~0.001.

We now summarize in Fig. 2e the drift coefficient ν of a-$CrTe_3$ (red stars) together with various amorphous PCMs[23-29,37-39] , plotted versus holding temperature. Previous drift measurements of a-PCMs, including the best-so-far $TiTe_2/Sb_2Te_3$ heterostructure [26], were mostly conducted at RT. The holding $T$ was limited to ~100 ºC, due to their relatively low $T_x$. Doping and alloying can increase the thermal stability of the amorphous phase, which, however, can also result in resistance drift in the crystalline phase with ν reaching ~0.01 or even higher [39]. By contrast, for our c-$CrTe_3$ (Extended Data Fig. 3), the sheet resistance showed a consistently low ν ~0.001 at all the holding temperatures used for a-$CrTe_3$. Scaling the film thickness down to several nm in general helps to reduce ν at RT [37], but it would be difficult to generate a large number of intermediate states accurately in such ultrathin films due to the limited active programming volume. We also note that $GeTe/Sb_2Te_3$ and $Sb_2Te_3$/Ge-Sb-Te superlattices may display a relatively low drift with ν~0.002−0.009 [40-42]. Nevertheless, their RESET states may be in a partly crystalline state. As seen in Fig. 2e, our $CrTe_3$ is the only PCM that is immune to resistance drift at all practical operating temperatures for IMC applications.

**Electrical measurement of $CrTe_3$ devices**

To ascertain the melt-quenched a-$CrTe_3$ displays the same no-drift behavior, we carried out electrical measurements on a series of confined memory devices, inside which a $CrTe_3$ thin film of ~100 nm is confined by the electrodes (tungsten) and the dielectric layers ($SiO_2$), as sketched in Fig. 3a. Further details about the device structure can be found in Supplementary Fig. S7. Firstly, the $CrTe_3$ cell was thermally annealed at 350 ºC to form a full crystalline state, and was programmed to a full amorphous state using a single electrical RESET pulse (Fig. 3b). The drift coefficient (ν ~0.001) determined from the data acquired immediately after the electrical melt-quench (see also the Supplementary Video 1) is consistent with the behavior of as-deposited amorphous phase. The subsequent cross-sectional TEM characterizations confirmed that the electrical switching rendered the cell fully amorphous (Fig. 3c). We also performed cycling measurements and found the same no-drift behavior of the RESET state of a-$CrTe_3$ after $2\times10^5$ cycles (Fig. 3d and 3e). The same electrical measurements were repeated for several additional cells, and the drift coefficient values of both the initial RESET and the final RESET states are consistently close to 0.001 (Extended Data Fig. 2). Meanwhile, the SET state (i.e., c-$CrTe_3$) showed constant electrical resistance over time as well (Fig. 3f). Further cycling experiment and cross-sectional TEM analysis are documented in Supplementary Fig. S8. We also note that the programming noise of the $CrTe_3$ device can be suppressed by increasing the pulsing width to reduce the crystallization randomness (Supplementary Fig. S9). Besides, we confirmed that consistently high resistance drift would persist for GST devices after extended cycling (Supplementary Fig. S10).

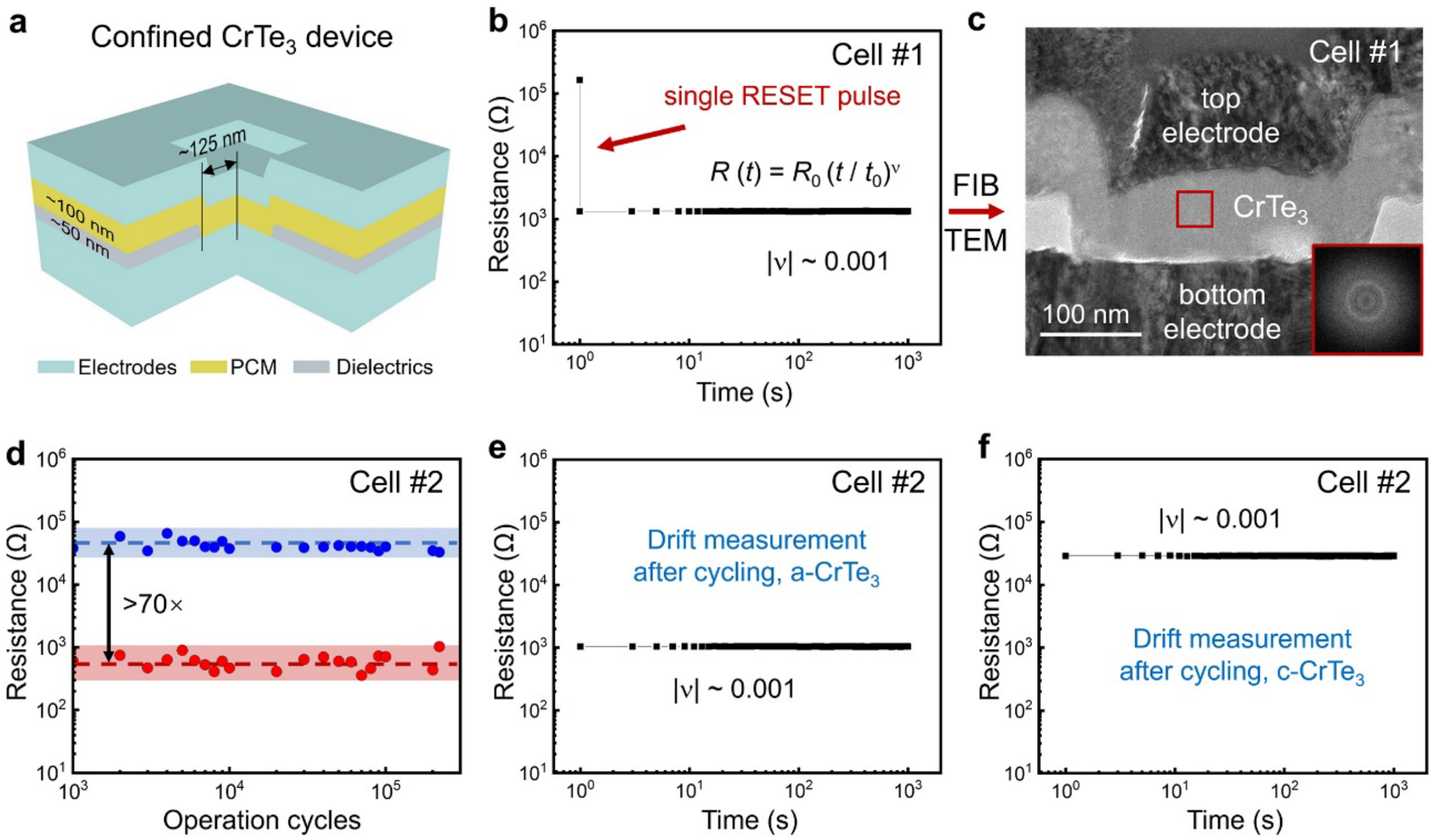

**Fig. 3 | Electrical measurements using confined $CrTe_3$ devices.** (**a**) The sketch of the confined electronic device with $CrTe_3$ being the functional PCM layer. (**b**) A single electrical pulse was applied to fully RESET a confined memory cell, after which its resistance was monitored immediately at RT. (**c**) The subsequent cross-sectional TEM characterizations of the RESET state, proving that the cell turned fully amorphous (see halos in the FFT pattern). (**d**) The electrically switched cycling test of a second confined cell. The drift measurement at RT of (**e**) a full RESET state and (**f**) a full SET state after cycling.

The essence of multilevel programming in PCM devices is to generate a series of intermediate resistance states, each corresponding to a different ratio of amorphous to crystalline volume. However, fine-tuning the volume fraction of crystalline versus amorphous regions using electrical pulses requires sophisticated device fabrication and control, and is subject to much uncertainty. We therefore developed a hybrid opto-electro-thermal scheme, which serves the purpose of assessing the multilevel capacity of $CrTe_3$ in a deterministic way. The precise volume control is achieved via step-wise SET laser operations, i.e., generating crystalline domains of the same size, one by one in a regular pattern, via laser irradiation on an amorphous thin film. The actual image of the measurement setups is displayed in Supplementary Fig. S9. Using a bridge-like device covered with as-deposited a-$CrTe_3$ thin film, we applied laser pulses in the middle part of devices for a sequential crystallization along the lateral direction (see the optical images in Supplementary Fig. S10). After sending every two laser pulses, the electrical resistance was measured and monitored for over 1 hour. During the whole process (> 16 hours), the three devices were each held at 25 °C, 80 °C or 150 °C, respectively. Each device displayed 16 well-separated resistance levels (Extended Data Fig. 5). The resistance levels can still be well distinguished even at the holding temperature of 165 °C (Supplementary Fig. S11).

Although crystallization rate is not a top-priority parameter to optimize owing to the parallel programming ability of PCM array [43], it is still desirable to have relatively fast crystallization to shorten the training time of the neural network. For the above operation, we used laser pulse of 100 ns (90 mW) to generate crystallized a-$CrTe_3$ domains, as confirmed by Raman spectroscopy measurements in Extended Data Fig. 4a. This crystallization speed may seem surprisingly fast, considering the a-$CrTe_3$ is rather stable against crystallization with a high activation energy for crystallization, $E_a$ ~3.28 eV (Extended Data Fig. 4b, higher than the 2.2 eV known for GST). This is because fast crystal growth in $CrTe_3$ could proceed via relatively quick rearrangement of [$CrTe_6$] octahedra, given sufficient thermal energy, as the Cr and Te atoms still move collectively, i.e., the robust octahedra do not fall apart into individual atoms that diffuse independently (see AIMD calculations at 300 and 400 °C in Extended Data Fig. 4c).

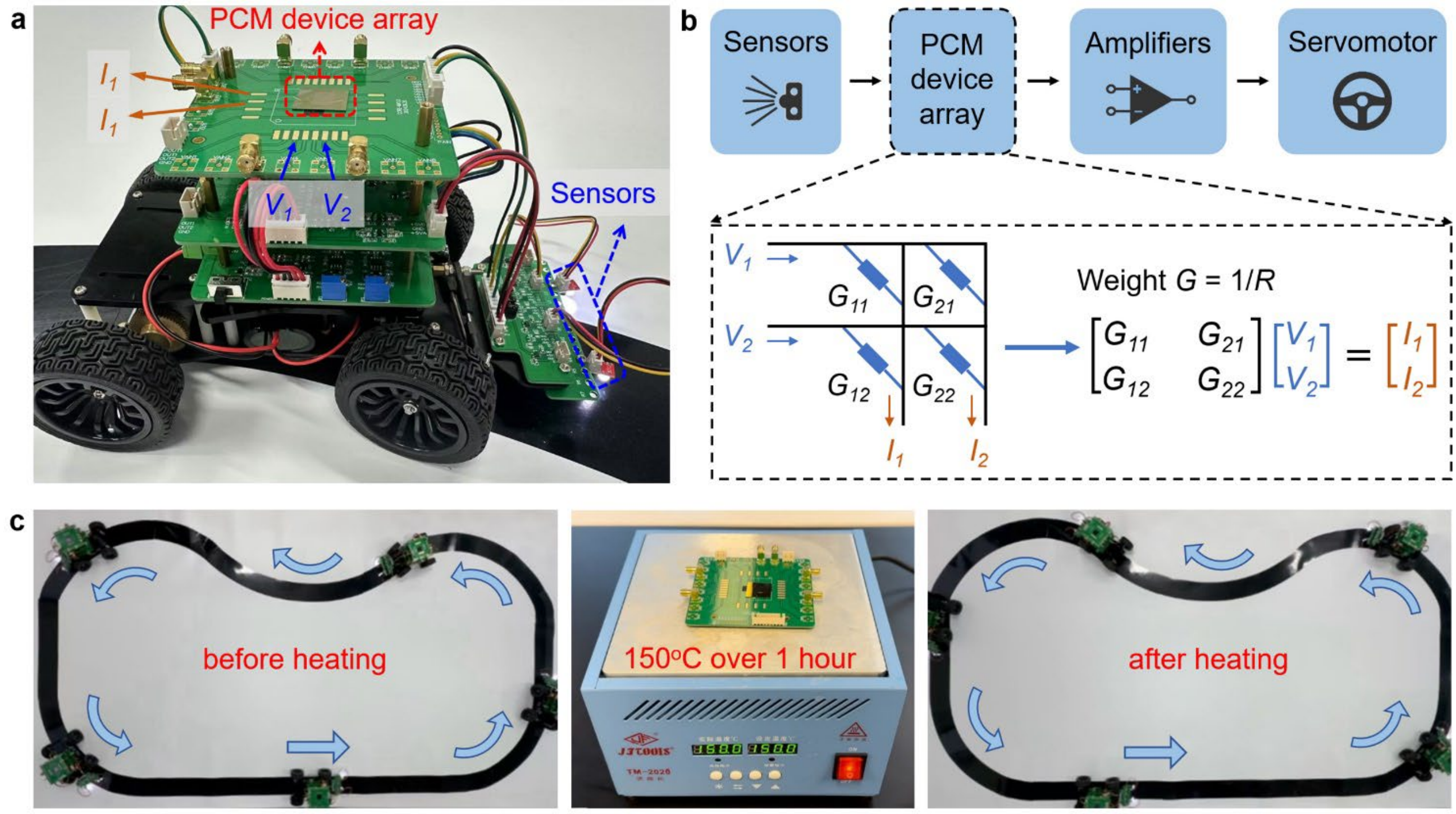

**Fig. 4 | The $CrTe_3$-based path-tracking vehicle.** (a) The image of $CrTe_3$-based path-tracking vehicle. (b) The schematic of the control system with a 2×2 $CrTe_3$ device array. (c) The path-tracking function was preserved after heating. See Supplementary Video 2 for the entire test process.

**Path-tracking function enabled by stable resistance levels**

To highlight the importance of robust resistance levels without resistance drift, we fabricated a 2×2 array of $CrTe_3$-based bridge-like devices, and integrated the array into a self-made vehicle (Fig. 4a) to realize stable path-tracking function. As shown in Fig. 4b, the control system of this vehicle mainly composed of two grayscale sensors, the device array, the amplifiers and the servomotor. The sensors are symmetrically installed in the front of the vehicle to detect grayscale changes, which provide distinct input voltage signals when detecting black and white. We programmed the 4 well-separated resistance values into the 4 devices via step-wise SET laser operations on as-deposited a-$CrTe_3$ thin films. The weight (conductance) values of the 4 devices were then mapped into a single-layer neural network (the 2×2 array) as the basic neuromorphic computing unit, performing matrix vector multiplication through Ohm's law and Kirchhoff's law to enable automatic path-tracking [44]. The output current signals from the array were then processed through the transimpedance amplifier and differential amplifier, and were converted into voltage signals again for the final steering and turning actions of the servomotor. The threshold for failure is set as tight as “any weight value changing by more than ±1%”. As shown in Fig. 4c and the Supplementary Video 2, the vehicle tracked the black path nicely, consistently turning the wheels back to the black path when detecting the white background. This $CrTe_3$-based path-tracking vehicle still functioned very well after being placed in air for over 1 month, and after heating the control board at 150 ºC for over 1 hour. As a control experiment, we have prepared a 2×2 array of GST-based bridge-like devices, and implemented it into the same vehicle. The automatic path-tracking function was achieved initially, but got completely lost after only 10 minutes (Supplementary Fig. S14 and Supplementary Video 3) due to the resistance drift of a-GST.

**Conclusions and outlook**

We have designed an unconventional PCM without resistance drift at almost all practical IMC operating temperatures ($T$=−200 ºC to 165ºC) from the get-go (immediately after the amorphous state was generated) to many hours of aging at each $T$. The materials science origin of this unprecedented behavior can be summarized as follows. The new a-PCM is characterized by “molecule-like” motifs that are unique in configuration, overwhelmingly predominant throughout the structure, and robust over time. To explore the general validity of such a structural feature, we carried out AIMD simulations and electrical measurement of amorphous $TiTe_2$ and $VTe_2$. As shown in Supplementary Fig. S12, the molecule-like motifs of [$TiTe_6$] and [$VTe_6$] octahedra also prevailed, and the measured drift coefficient of a-$TiTe_2$ and a-$VTe_2$ thin films was as small as that of a-$CrTe_3$. Nevertheless, the limited contrast window and the low crystallization temperature of the two alloys render them unfit for practical applications. We also note that not all Te-rich transition metal tellurides should be regarded as molecule-like PCMs. In a very recent work [45], a low-drift behavior was found for a-$NbTe_4$ thin film at room temperature. But no clear molecule-like pattern was observed in the a-$NbTe_4$ amorphous model obtained via AIMD simulations. Indeed, this low-drift behavior cannot be sustained at slightly elevated temperatures: the drift coefficient quickly increased to ~0.082 at 80 ºC (see Supplementary Fig. S13).

Overall, our work represents a different avenue towards ultralow resistance drift, by resorting to a glass with molecule-like motifs. In contrast with conventional a-PCMs (or other glass categories) that tend to readily relax towards other adjacent sub-basins (variable motif configurations) across a fairly shallow megabasin in the potential energy landscape, the “molecular-glass-like” a-PCM we designed would stay put in a rather steep basin (see schematic in Extended Data Fig. 6), with little motivation for relaxation and no other desirable motifs that can be accessed (at least up to ~200°C). The transformation between amorphous and crystalline $CrTe_3$ via the cooperation of compact and robust octahedra also resulted in a desirable combination of excellent amorphous stability and satisfactorily fast crystallization speed. Future work is anticipated to gain a deeper understanding of this alloy at the temperatures approaching crystallization via *in situ* measurements [46,47] and machine-learning-facilitated molecular dynamics [48]. Also, it is necessary to test the multilevel capacity in nanoscale PCM devices [49], the endurance beyond million cycles, as well as the device variability of $CrTe_3$ in a large crossbar array, before the new PCM can be put into practical use. Nevertheless, we have demonstrated that $CrTe_3$ thin films can be produced with wafer-scale homogeneity via a standard sputtering approach, adding no complexity to mass production nor to programming algorithm. Overall, our no-drift discovery achieves the first intrinsically robust a-PCM, offering a front-end materials solution that has the potential to meet the demanding requirements of high-precision multilevel programming at the operation temperatures for both data-centric and edge-computing platforms.

**ACKNOWLEDGMENTS**
The work is supported by the National Key Research and Development Program of China (2023YFB4404500). W.Z. thanks the support of National Natural Science Foundation of China (62374131). J.-J.W. thanks the support of National Natural Science Foundation of China (62204201). We acknowledge Dr. Xi Li for useful discussions. We thank Chao Li and Chuansheng Ma for their helps on TEM characterizations, Songquan Yang for his technical support on bridge-like device fabrication, and Weiye Liu for his assistance with the integration of the path-tracking vehicle. We acknowledge the HPC platform of XJTU and the Computing Center in Xi'an for providing computational resources. The authors acknowledge XJTU for hosting their work at CAID. The International Joint Laboratory for Micro/Nano Manufacturing and Measurement Technologies of XJTU is acknowledged.

**Author contributions:**
W.Z., X.W., and E.M. conceived the idea and designed the experiments and simulations. X.W., J.-J.W. and R.W. carried out most of the experiments with the help of D.X., C.N., Z.Z., C.W., J.Z., W. Zhou., and Z.S. S.S. performed DFT calculations with the help of R.C. and X.S. D.X. and J.-J.W. did the image recognition simulations and the vehicle testing. W.Z., X.W. and E.M. wrote the manuscript with input from J.-J.W. and S.S. All authors discussed the results and approved the submission of the manuscript.

**Competing interests:**
The authors declare no competing interests.

**Data and materials availability:**

All data needed to evaluate the conclusions in this paper are included in the paper or the supplementary materials. Several patents (PCT/CN2024/123713, CN 202411271055.6, CN 202411271061.1) on no-drift PCM are under examination.

# Methods

## Synthesis and structural characterizations

The $CrTe_3$ thin films of ~50−150 nm were deposited on 4-inch $SiO_2$/Si substrates at room temperature with a pure Cr target and a pure Te target in high vacuum via AJA ORION 8 (base pressure of less than ~$1\times10^{-5}$ Pa). The deposition rate was set to 9.5 nm $min^{-1}$. The chemical composition of the sputtered films was determined to be approximately $Cr_{25.4}Te_{74.6}$ by the energy-dispersive X-ray (EDX) experiments. The structures of the as-deposited and post-annealed thin films were investigated by XRD with Cu Kα radiation (Bruker D8 ADVANCE). The TEM specimens were prepared by a dual beam FIB system (Helios NanoLab 600i, FEI) with a Ga ion beam operated at 30 kV. A Pt protective layer was deposited above the $CrTe_3$ thin film to mitigate potential damages from Ga ions during the FIB lift-out and thinning processes. The TEM experiments were performed on a Talos-F200X operated at 200 kV. The STEM−HAADF imaging and EDX mapping experiments were performed on a JEOL ARM200F STEM with a probe aberration corrector, operated at 200 kV. Raman spectra were collected using Renishaw inVia Qontor with a solid-state 532 nm laser for the excitation, where the laser power was set as 0.25 mW, and an exposure time of 2 seconds with 50 cycles was used.

## Fabrication of devices and the vehicle

The confined memory devices were fabricated using electron beam lithography and reactive ion etching. The top and bottom electrodes were made of tungsten and $SiO_2$ was used for the dielectric layers. The thickness of the $CrTe_3$ layer was ~100 nm. Two lithography processes were adopted for the fabrication of the bridge-like devices. The first lithography process was applied to pattern the layer of metal electrode on a $SiO_2$ wafer using a UV lithography machine (SUSS MJB4). The tungsten electrodes of 30 nm thickness were deposited on the patterned substrate via magnetron sputtering followed by lift-out process. The second lithography process was applied to pattern the PCM layer. The deposited $CrTe_3$ layer (~50 nm) was slightly thicker than that of the tungsten electrodes in order to ensure the connectivity of the device. A ZnS-$SiO_2$ protective layer with a thickness of 20 nm was deposited on top of the surface. Both lithography processes were done using a negative photoresist (AZ5214E). The 2×2 $CrTe_3$ device array was fabricated and integrated on a custom-printed circuit board (PCB) using a wire bonder (TPT HB10). The vehicle consisted of two grayscale sensors, a $CrTe_3$ array, amplifiers and a motor system (including servomotor and engine), batteries and the vehicle body. The servomotor performed different steering actions directed by the $CrTe_3$-based control system.

## Electrical measurements

The home-made electrical testing system was constructed by connecting the Keithley 2636B source meter and the Instec mK200 hot stage with a temperature accuracy of 0.001 °C. The sheet resistance of the $CrTe_3$ films as a function of temperature was measured with a heating rate of 10 °C / min under Ar protection. The probe electrodes were made of tungsten. For the electrical measurements of 34 units on the wafer, we first broke down these units from the wafer, and then repeated the sheet resistance measurement on each of them. The PPMS instrument (Quantum Design DynaCool) was used for the Hall effect measurement at room temperature, and the low-temperature resistance measurements between −50 to −200 ºC. The electrical measurements of confined memory devices were performed using the Keithley 2400C source meter (measuring cell resistance) and the Tektronix AWG5002B pulse generator (generating voltage pulse). Long voltage pules were employed in the cycling experiments to guarantee the devices to form full RESET (2.5 V, 300 ns followed by a 40 ns falling edge) and SET states (1.5 V, 500 ns followed by an 80 ns falling edge) for the subsequent resistance drift measurements. The self-built opto-electro-thermal platform was equipped with Stradus 488-25 lasers, arbitrary function generator (AFG 31000 SERIES), the optical path system, Keithley 2636B source meter, Instec mK200 hot stage and a computer. The wavelength of the applied laser was fixed at 488 nm, and the pulse duration and power were set as 100 ns and 90 mW for step-wise crystallization of the bridge-like devices. The source meter records the resistance values continuously so that any change in resistance can be monitored in real time upon multilevel programming. Slight numerical difference in the electrical resistance value of the amorphous (or crystalline) state is expected for different device setups, in which the film thickness, length, type of electrodes, amorphous-to-crystal ratio, orientation of the crystal, etc., can lead to some resistance variations.

**Ab initio calculations**

We carried out DFT-based AIMD simulations using the second-generation Car–Parrinello molecular dynamics scheme [46] as implemented in the CP2K package [47]. The Kohn–Sham orbitals were expanded by a double- and triple-zeta plus polarization Gaussian-type basis set for Cr and Te, and the charge density was expanded in plane waves with a cutoff of 300 Ry. The Goedecker–Teter–Hutter (GTH) pseudopotentials [48] and Perdew–Burke–Ernzerhof (PBE) functional [49] with van der Waals (vdW) correction based on the Grimme's D3 method [50] were applied, and the spin-polarization was considered using α and β orbitals without spin restriction. The AIMD calculations were carried out using the canonical ensemble (NVT) with a time step of 2 fs. To generate a-$CrTe_3$, 50 Cr atoms and 150 Te atoms were randomly arranged in a cubic cell. The model was heated above 3000 K to remove all possible crystalline order, and were then quenched down to ~1200 K in 10 ps. After an AIMD run at ~1200K for 30 ps, the model was then quenched down to 0 K with a cooling rate of 12.5 K $ps^{-1}$. To collect structural data at different holding temperatures, the corresponding snapshots within the cooling process were picked out for a separate 50-ps AIMD run at the respective holding temperature. Three amorphous models with independent thermal history were calculated in both crystalline and amorphous densities, respectively. The structural relaxation and further spin-polarized density of states analyses at 0 K were conducted using the VASP code [51] with PBE functional, vdW-D3 correction, and the projector augmented-wave (PAW) pseudopotentials [52]. The energy cutoff was set as 450 eV. Each amorphous model contained 50 Cr atoms and 150 Te atoms, and its Brillouin zone was sampled using *Γ* point. For c-$CrTe_3$, the unit cell contained 8 Cr atoms and 24 Te atoms, and its Brillouin zone was sampled using a 2×2×4 k-point mesh.

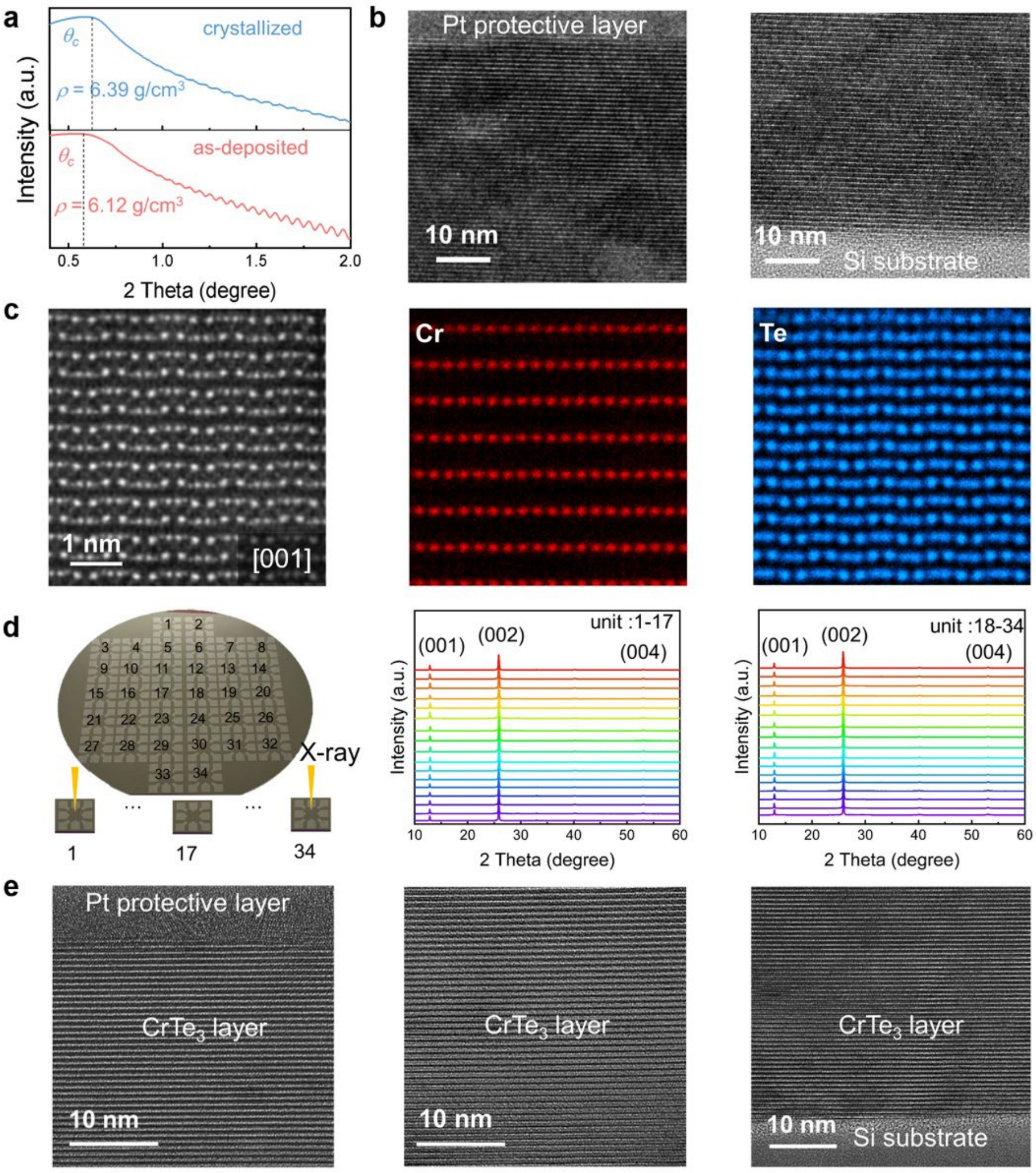

**Extended Data Fig. 1 | Structural characterizations of $CrTe_3$ thin films.** (a) The XRR spectrum of as-deposited amorphous films and crystallized $CrTe_3$ films of ~150 nm. (b) The TEM images near top Pt protective layer and the Si substrate of crystallized $CrTe_3$ film of ~150 nm. (c) The atomic-scale HAADF image of crystalline $CrTe_3$ viewed in the [001] direction and its corresponding EDS mapping of Cr and Te. The intensity of spots in the HAADF image is approximately proportional to the square of the averaged atomic number Z of each column along the view direction, therefore, the Te (Z = 54) columns look brighter than the Cr (Z = 24) columns. (d) The measured XRD patterns over 34 locations on the 4-inch wafer. (e) The TEM images recorded at different locations of crystallized $CrTe_3$ film of ~50 nm. No clear grain boundary can be observed over >5 μm lateral direction. The film thickness along the beam incidence direction was ~80 nm for all TEM specimens we prepared.

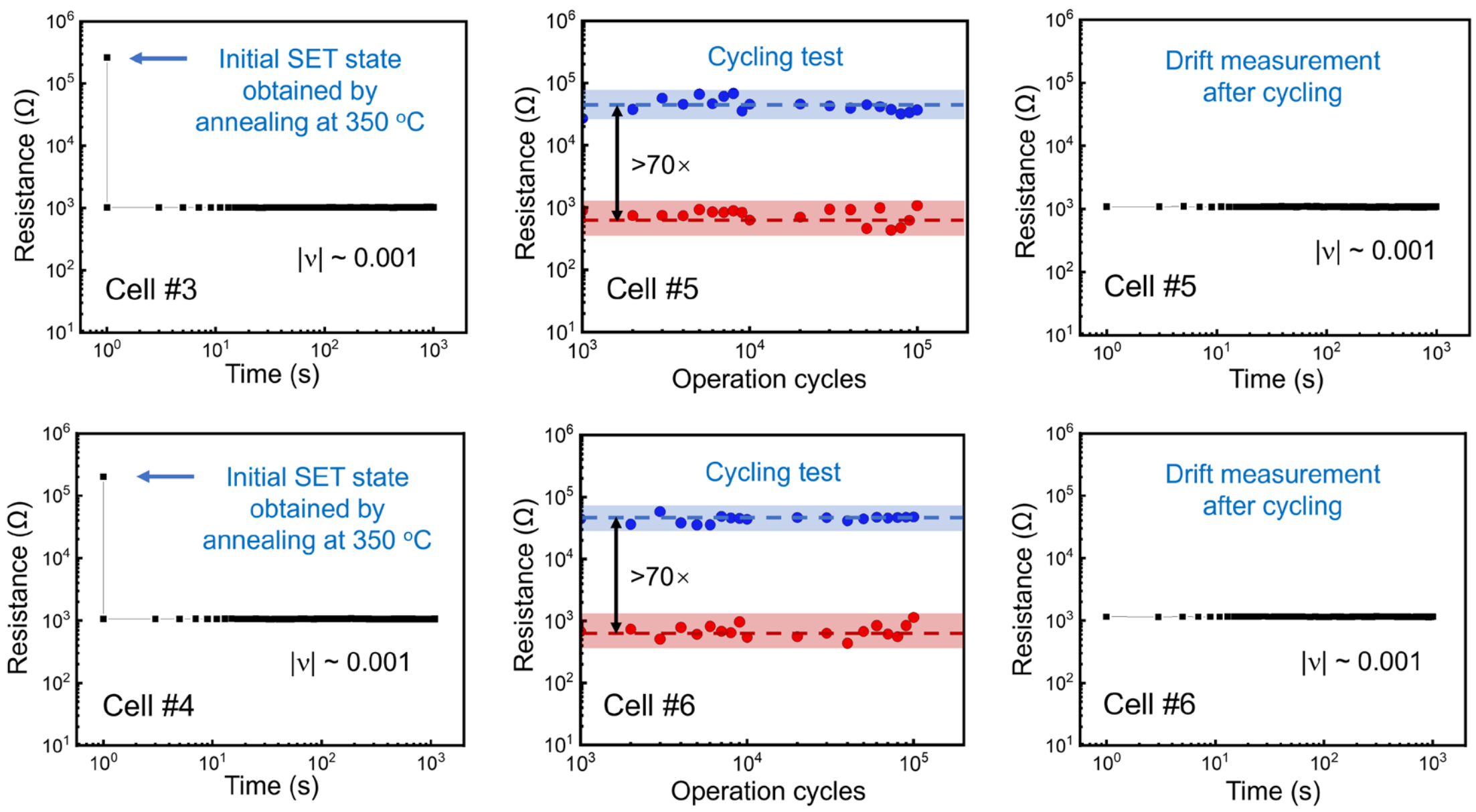

**Extended Data Fig. 2 | The drift measurements after the initial electrical melt-quench RESET operation in electronic device, the cycling endurance test itself, and the drift measurements of the full RESET state after cycling at RT.** Electrical measurements of four additional confined memory cells confirmed the no-drift behavior of the RESET state. See Supplementary fig. S9 for the effect of pulse width in suppressing programming noise.

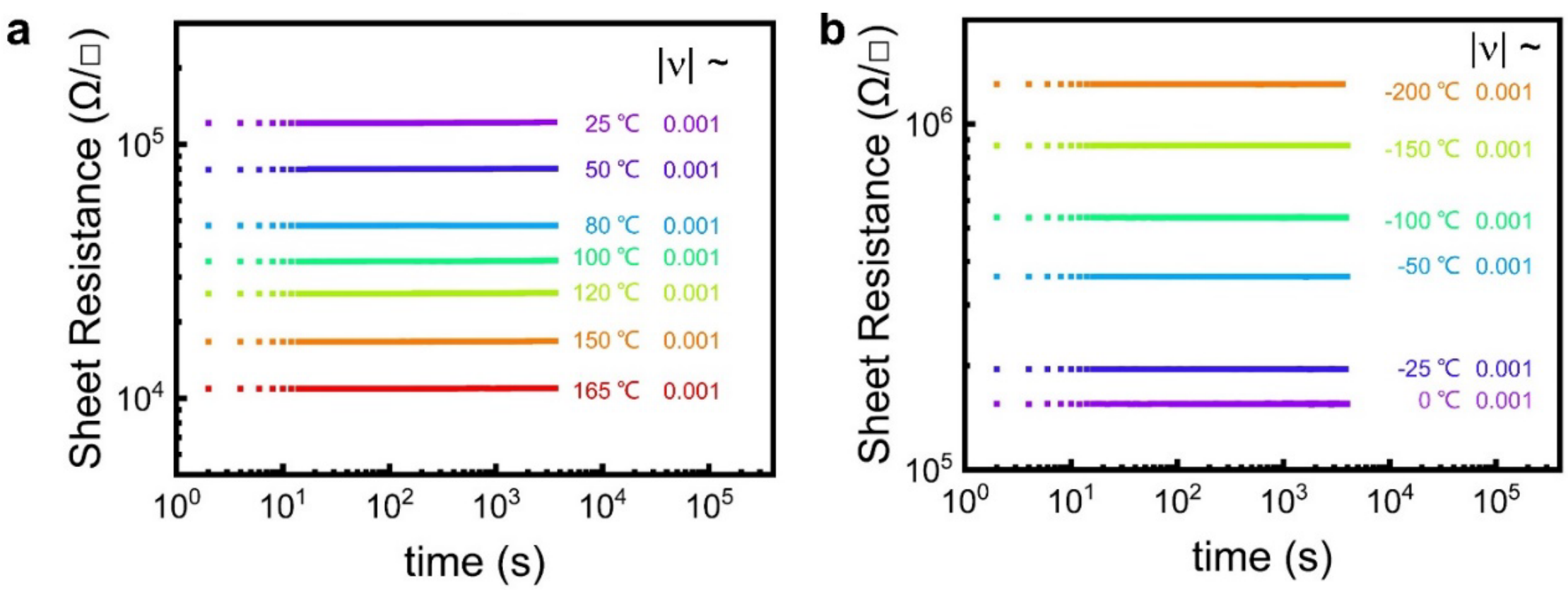

**Extended Data Fig. 3 | The resistance drift measurements of the crystallized $CrTe_3$ thin films, each conducted at a different holding temperature** (a) at or above RT, and (b) at 0 °C or below.

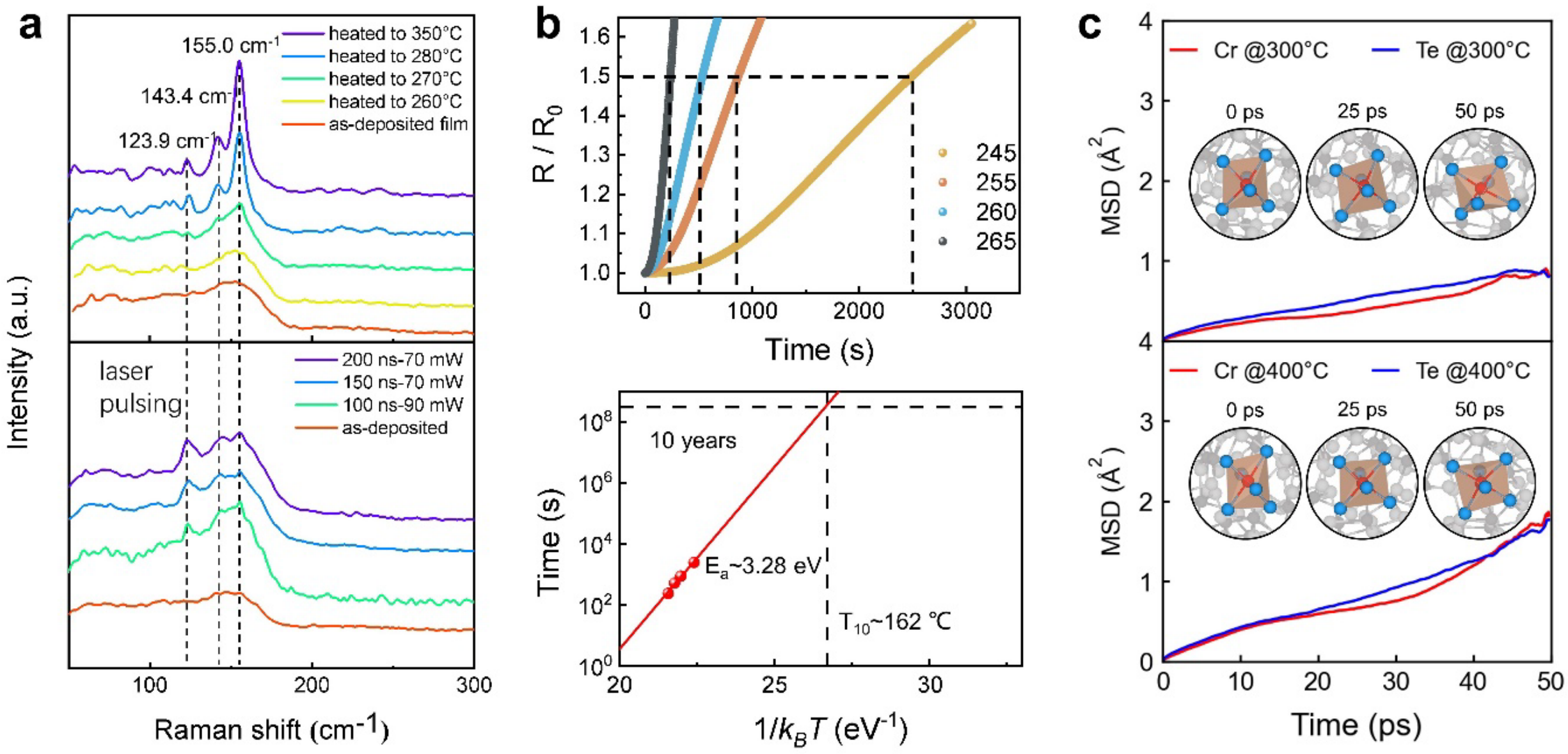

**Extended Data Fig. 4| Switching and dynamics.** (a) The measured Raman spectrum of the as-deposited film, and the films being heated to 260 °C, 270 °C, 280 °C and 350 °C, respectively (upper panel). The Raman spectrum measured at the laser-irradiated spots of the films (lower panel). (b) The electrical resistance of as-deposited $CrTe_3$ films measured at 245 ºC, 255 ºC, 260 ºC and 265 ºC, respectively (upper panel). The extrapolated 10-year data retention temperature $T_{10\text{-year}}$ and the estimated activation energy $E_a$ of crystallization for a-$CrTe_3$ (lower panel). (c) The mean squared displacement (MSD) curves of a-$CrTe_3$ at 300°C and 400°C via AIMD calculations. The insets show a series of snapshots of a typical Cr[$Te_6$] octahedron taken at 0, 25 and 50 ps of the AIMD trajectories at respective temperatures. The Cr[$Te_6$] octahedron was robust, and the associated 1 Cr atom and 6 Te atoms moved collectively. At both holding temperatures, the fraction of Cr atoms in Cr[$Te_6$] octahedron is over 90% despite of thermal fluctuations.

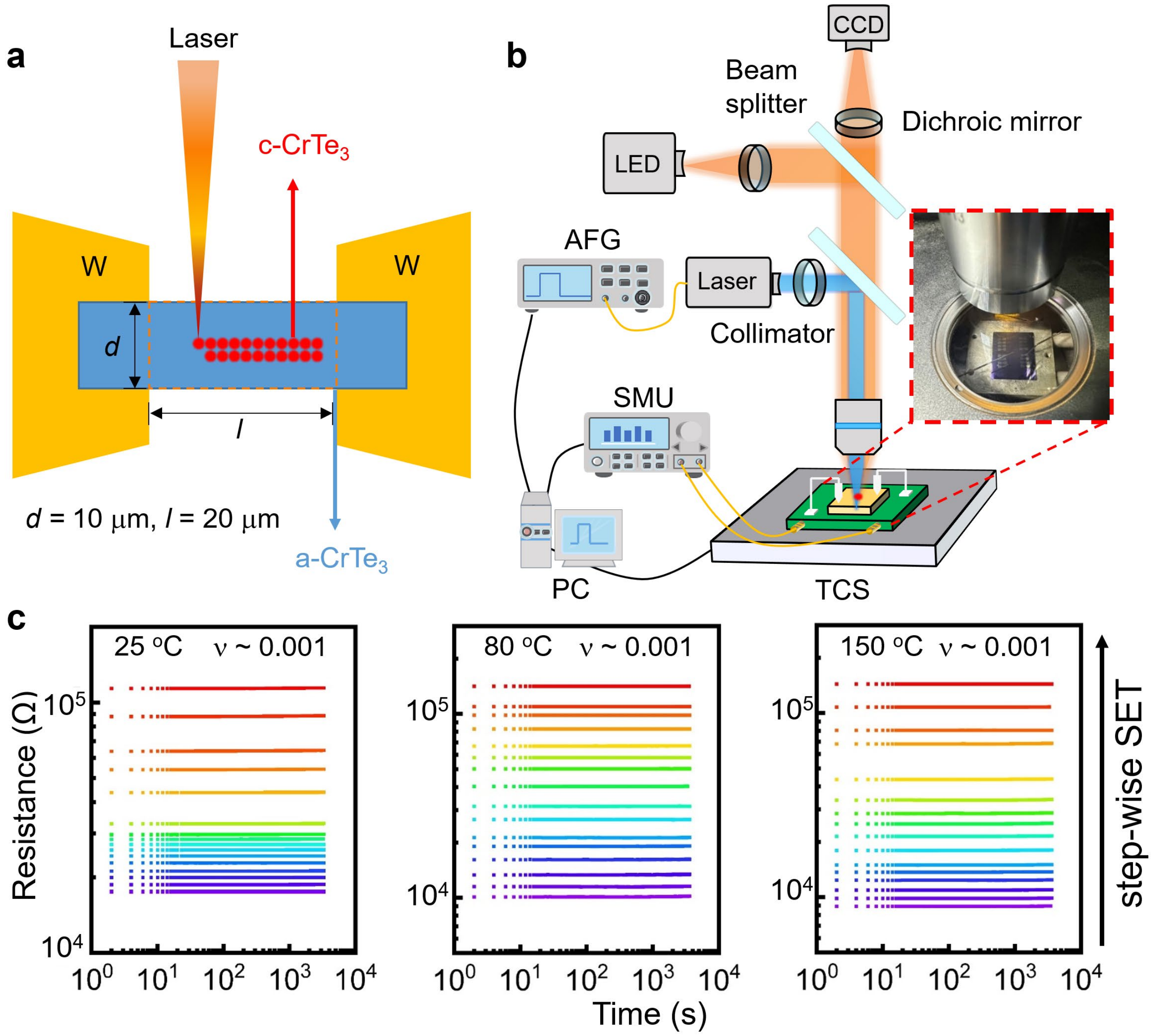

**Extended Data Fig. 5| Deterministic multilevel programming.** (a) The scheme of optical programming and electrical measurement on a bridge-like $CrTe_3$ device. (b) The hybrid opto-electro-thermal setup consisted of a nanosecond laser beam with 1 μm spot size (controlled using an arbitrary functional generator, AFG), a source measure unit (SMU) for electrical measurements, and a temperature controller stage (TCS). The bottom-right inset shows the photo of the self-made $CrTe_3$ device. (c) Using a bridge-like device of 10×20 $\mu m^2$, 16 robust resistance states were obtained at 25, 80 and 150 °C via step-wise SET operations. Each resistance level was measured after sending two laser pulses.

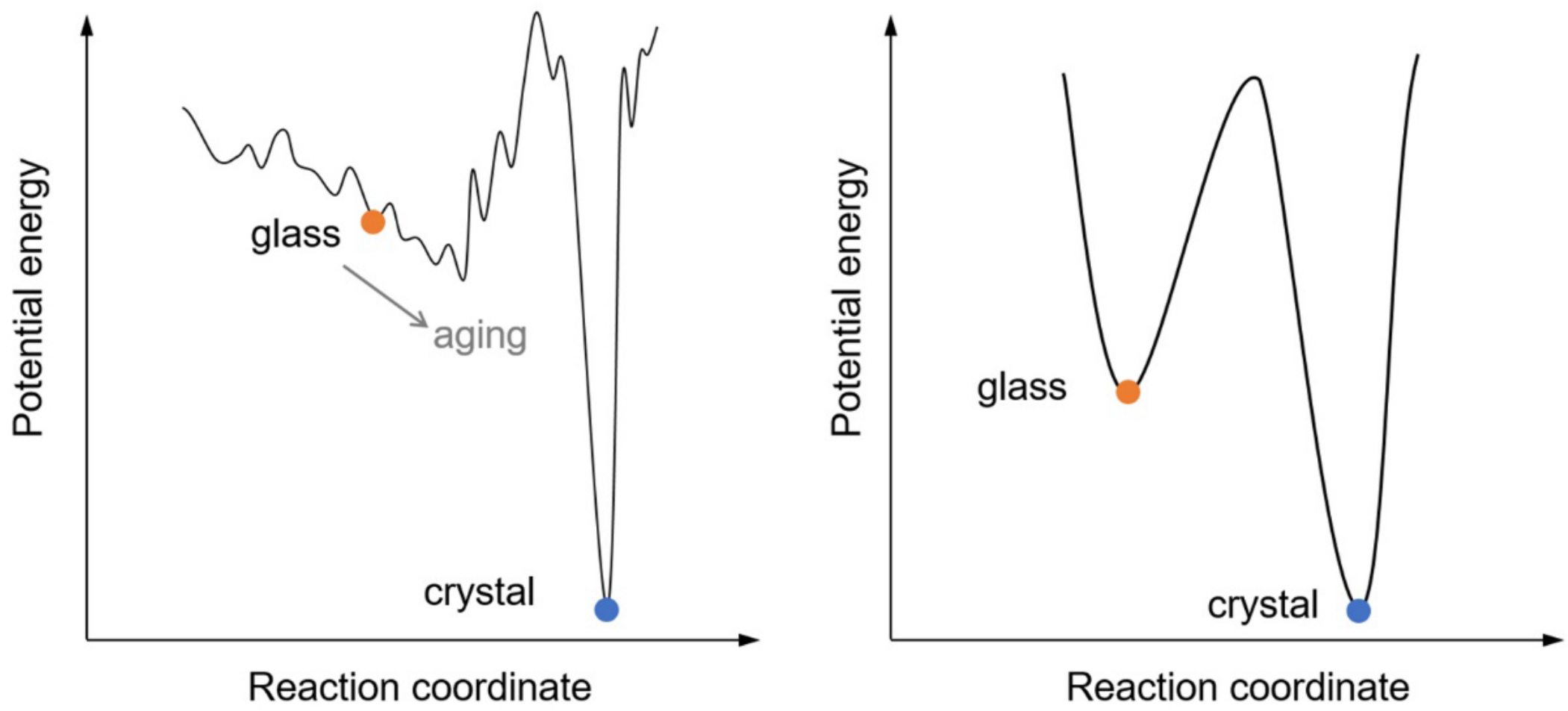

**Extended Data Fig. 6| Schematic of the potential energy landscape.** A substantial difference is expected for conventional PCMs (left panel, where upon ageing the glass can relax towards a number of possible lower-energy basins) and our new molecule-like PCMs (right panel, in which the glass sits at the bottom of a rather deep basin, while comparable/competing metastable basins are either non-existent, or all out of the picture because they are inaccessible due to high barriers for the glass aged in this study at temperatures below ~200ºC). See text for additional explanation.